\documentclass[11pt]{article}
\usepackage{amsmath,amsthm,latexsym,amssymb,amsfonts,epsfig,psfrag}

\makeatletter \@addtoreset{equation}{section} \makeatother

\def\be{\begin{equation}}
\def\ee{\end{equation}}
\def\ba{\begin{eqnarray}}
\def\ea{\end{eqnarray}}

\newcommand\nn{\nonumber}
\newcommand\q{\quad}
\def\Nl{{\mathchoice
{\setbox0=\hbox{$\displaystyle\rm N$}\hbox{\hbox to0pt
{\kern0.4\wd0\vrule height0.9\ht0\hss}\box0}}
{\setbox0=\hbox{$\textstyle\rm N$}\hbox{\hbox to0pt
{\kern0.4\wd0\vrule height0.9\ht0\hss}\box0}}
{\setbox0=\hbox{$\scriptstyle\rm N$}\hbox{\hbox to0pt
{\kern0.4\wd0\vrule height0.9\ht0\hss}\box0}}
{\setbox0=\hbox{$\scriptscriptstyle\rm N$}\hbox{\hbox to0pt
{\kern0.4\wd0\vrule height0.9\ht0\hss}\box0}}}}
\def\Zl{{\mathchoice
{\setbox0=\hbox{$\displaystyle\rm Z$}\hbox{\hbox to0pt
{\kern0.4\wd0\vrule height0.9\ht0\hss}\box0}}
{\setbox0=\hbox{$\textstyle\rm Z$}\hbox{\hbox to0pt
{\kern0.4\wd0\vrule height0.9\ht0\hss}\box0}}
{\setbox0=\hbox{$\scriptstyle\rm Z$}\hbox{\hbox to0pt
{\kern0.4\wd0\vrule height0.9\ht0\hss}\box0}}
{\setbox0=\hbox{$\scriptscriptstyle\rm Z$}\hbox{\hbox to0pt
{\kern0.4\wd0\vrule height0.9\ht0\hss}\box0}}}}
\def\Ql{{\mathchoice
{\setbox0=\hbox{$\displaystyle\rm Q$}\hbox{\hbox to0pt
{\kern0.4\wd0\vrule height0.9\ht0\hss}\box0}}
{\setbox0=\hbox{$\textstyle\rm Q$}\hbox{\hbox to0pt
{\kern0.4\wd0\vrule height0.9\ht0\hss}\box0}}
{\setbox0=\hbox{$\scriptstyle\rm Q$}\hbox{\hbox to0pt
{\kern0.4\wd0\vrule height0.9\ht0\hss}\box0}}
{\setbox0=\hbox{$\scriptscriptstyle\rm Q$}\hbox{\hbox to0pt
{\kern0.4\wd0\vrule height0.9\ht0\hss}\box0}}}}
\def\Rl{{\mathchoice
{\setbox0=\hbox{$\displaystyle\rm R$}\hbox{\hbox to0pt
{\kern0.4\wd0\vrule height0.9\ht0\hss}\box0}}
{\setbox0=\hbox{$\textstyle\rm R$}\hbox{\hbox to0pt
{\kern0.4\wd0\vrule height0.9\ht0\hss}\box0}}
{\setbox0=\hbox{$\scriptstyle\rm R$}\hbox{\hbox to0pt
{\kern0.4\wd0\vrule height0.9\ht0\hss}\box0}}
{\setbox0=\hbox{$\scriptscriptstyle\rm R$}\hbox{\hbox to0pt
{\kern0.4\wd0\vrule height0.9\ht0\hss}\box0}}}}
\def\Cl{{\mathchoice
{\setbox0=\hbox{$\displaystyle\rm C$}\hbox{\hbox to0pt
{\kern0.4\wd0\vrule height0.9\ht0\hss}\box0}}
{\setbox0=\hbox{$\textstyle\rm C$}\hbox{\hbox to0pt
{\kern0.4\wd0\vrule height0.9\ht0\hss}\box0}}
{\setbox0=\hbox{$\scriptstyle\rm C$}\hbox{\hbox to0pt
{\kern0.4\wd0\vrule height0.9\ht0\hss}\box0}}
{\setbox0=\hbox{$\scriptscriptstyle\rm C$}\hbox{\hbox to0pt
{\kern0.4\wd0\vrule height0.9\ht0\hss}\box0}}}}
\def\Hl{{\mathchoice
{\setbox0=\hbox{$\displaystyle\rm H$}\hbox{\hbox to0pt
{\kern0.4\wd0\vrule height0.9\ht0\hss}\box0}}
{\setbox0=\hbox{$\textstyle\rm H$}\hbox{\hbox to0pt
{\kern0.4\wd0\vrule height0.9\ht0\hss}\box0}}
{\setbox0=\hbox{$\scriptstyle\rm H$}\hbox{\hbox to0pt
{\kern0.4\wd0\vrule height0.9\ht0\hss}\box0}}
{\setbox0=\hbox{$\scriptscriptstyle\rm H$}\hbox{\hbox to0pt
{\kern0.4\wd0\vrule height0.9\ht0\hss}\box0}}}}
\def\Ol{{\mathchoice
{\setbox0=\hbox{$\displaystyle\rm O$}\hbox{\hbox to0pt
{\kern0.4\wd0\vrule height0.9\ht0\hss}\box0}}
{\setbox0=\hbox{$\textstyle\rm O$}\hbox{\hbox to0pt
{\kern0.4\wd0\vrule height0.9\ht0\hss}\box0}}
{\setbox0=\hbox{$\scriptstyle\rm O$}\hbox{\hbox to0pt
{\kern0.4\wd0\vrule height0.9\ht0\hss}\box0}}
{\setbox0=\hbox{$\scriptscriptstyle\rm O$}\hbox{\hbox to0pt
{\kern0.4\wd0\vrule height0.9\ht0\hss}\box0}}}}
\newcommand{\eps}{\epsilon}

\newcommand{\zp}{\,{}^0\!\phi}
\newcommand{\op}{\,{}^1\!\phi}
\renewcommand{\sp}{\,{}^2\!\phi}
\newcommand{\zl}{\,{}^0\!\lambda}
\newcommand{\ol}{\,{}^1\!\lambda}
\renewcommand{\sl}{\,{}^2\!\lambda}
\newcommand{\za}{\,{}^0\!\alpha}
\newcommand{\oa}{\,{}^1\!\alpha}
\newcommand{\sa}{\,{}^2\!\alpha}
\newcommand{\zx}{\,{}^0\!\xi}
\newcommand{\ox}{\,{}^1\!\xi}
\newcommand{\sx}{\,{}^2\!\xi}
\newcommand{\Zp}{ \Phi}

\newcommand{\Zx}{\Xi}

\title{ How to construct diffeomorphism symmetry on the lattice}
\author{ Bianca Dittrich\\
\small Perimeter Institute for Theoretical Physics,\\
\small 31 Caroline St. N, Waterloo, ON N2L 2Y5, Canada\\
\small and\\
\small   MPI for Gravitational Physics,\\
 \small Am M\"uhlenberg 1, D-14476 Potsdam, Germany \\
}

\date{}

\begin{document}

\maketitle

\begin{abstract}
\noindent
Diffeomorphism symmetry, the fundamental invariance of general relativity, is generically broken under discretization. After discussing the meaning and implications of diffeomorphism symmetry in the discrete, in particular for the continuum limit, we introduce a perturbative framework to construct discretizations with an exact notion of diffeomorphism symmetry. We will see that for such a perturbative framework consistency conditions need to be satisfied which enforce the preservation of the gauge symmetry to the perturbative order under discussion.  These consistency conditions will allow structural investigations of diffeomorphism invariant discretizations.

\end{abstract}


\section{Introduction and overview}

Lattice discretizations of field theories are a popular method to access non-perturbative quantum physics, for instance very successfully in lattice quantum chromodynamics. Similarly, many approaches to quantum gravity are based on discretizations \cite{lollreview}, such as (quantum) Regge calculus \cite{regge} or spin foams \cite{sf}. There is however, an important difference between the status of discretizations available for Yang Mills theories and for (4D) general relativity. Whereas for the former, discretizations are available that do preserve the Yang Mills gauge symmetry also on the lattice \cite{wilson_action} this is not the case for gravity \cite{bahrdittrich1,dittrichhoehn}, where the gauge symmetry in question is given by  diffeomorphism symmetry. The reason is that diffeomorphism symmetry acts on space time itself. If this space time is discretized, we can expect that a diffeomorphism would deform in some way this discretization. 

Indeed, there are several examples where diffeomorphism symmetry is realized also for the discretization (this includes 3D gravity \cite{louapre,improve}, reparametrization invariant 1D systems \cite{proc,seb} and linearized 4D gravity \cite{roc}), both at the classical and quantum level. For all these examples, diffeomorphism symmetry acts by displacing the vertices of the lattice in the space time in which the lattice is embedded. (This kind of diffeomorphism symmetry in the discrete was termed ditt--invariance in \cite{ditt}.) That is this symmetry can change the geometrical distance between the vertices. Here one can already see that such a discrete notion of diffeomorphism symmetry is enormously powerful: a  discrete system in which such a symmetry is realized needs to reproduce physics on all length scales, also on the large ones. The exploration of the consequences of such a symmetry has been only recently started, see for instance \cite{review,seb,ditt,sigmaben}. In this work we will see that on the one hand it is very complicated (or might not be possible) to construct discretizations with such a symmetry of a given system. We will therefore propose a perturbative approach. On the other hand such a symmetry has a number of important advantages and moreover would solve long standing problems for discretizations, in particular of gravity:
\begin{itemize}
\item
{\it Consistent perturbative formalism:} The main body of this paper will discuss how to obtain a consistent perturbative formalism for discretizations, in which a gauge symmetry is broken at a certain order (as is the case for 4D gravity). For instance Regge gravity will not allow a consistent perturbative framework around flat space, if one does not improve the action appropriately. The problem is, that linearized Regge gravity displays the linearized form of diffeomorphism symmetry, i.e. one can identify longitudinal lattice modes, that do not propagate as these are null modes of the Hessian of the action \cite{roc,he}. The higher order interactions will however involve these longitudinal modes. That is at higher order the gauge freedom associated to these modes does get fixed. This happens however in a non-linear fashion. Basically, the perturbation assumption, namely that the solution is analytical in a small parameter $\epsilon$ is not valid \cite{dittrichhoehn}. This means that for instance the computation of graviton scattering is not possible without changing the given discrete action to have a perturbative consistent form to the required order. As this problem is rooted in having broken symmetries, it will not appear if a discrete notion of diffeomorphism symmetry is exactly realized. On the other hand we will see that the requirement of perturbative consistency might help us to construct discretizations with such a symmetry.

\item
{\it Canonical formalism with first class constraints:} A long standing problem in discrete gravity is the construction of a consistent canonical formalism. In the continuum the dynamics in the canonical formalism is generated by arbitrary combinations of the Hamiltonian and (spatial) diffeomorphism constraints. The arbitrariness of the coefficients -- lapse and shift -- reflects the diffeomorphism symmetry of the covariant formalism and indeed the constraints do follow  from the diffeomorphism symmetry of the theory. Discretizations (in 4D) break this symmetry. Hence in the discrete, we cannot expect constraints and also not the related gauge freedom of freely choosing lapse and shift. Indeed these get rather fixed to some discrete values in the cases where the symmetries are broken. This also means that time evolution will proceed in discrete steps \cite{consistent,sigmaben}.    \\
This situation is not so much a problem in the classical realm. One can define a canonical formalism that exactly reproduces the solutions of the covariant one \cite{consistent, review, bahrdittrich1, dittrichhoehn, simplicial}, together with the exactly preserved and broken symmetries. Recently a canonical formalism has been defined which can handle arbitrary triangulations and the associated issue of changing phase space dimensions during time evolution \cite{simplicial}. It therefore can reproduce for instance all Regge solutions.\\ 
One has however to realize that having constraints in the continuum which are not reproduced in the discrete does lead to repercussions. Indeed constraints are just equations of motions, which involve the data of one time slice only. If the associated symmetry is broken by the discretization, this equation of motion will be a proper one, i.e. describing a coupling between time slices, however this coupling will be very weak. These equations are termed pseudo constraints, and can be imagined as describing thickened out constraint hypersurfaces \cite{bahrdittrich1}. The problem now is that selecting canonical data `far away' from this hypersurface will lead to unphysical solutions (not approximating a continuum solution), resulting for instance in complex lapse and shift parameters. Classically one could deal with this problem by staying near this pseudo constraint hypersurface. In quantum theory however it is unclear how to deal with such pseudo constraints. Proper (first class) constraints have to be imposed onto the quantum states, this is however not possible for the pseudo constraints which are not first class, i.e. do not form an algebra. This problem is addressed in the uniform discretization program \cite{uniform, sigmaben} in which all the (pseudo) constraints are squared and summed to one master constraint \cite{master} thus avoiding inconsistencies due to the constraint algebra. It is however unclear whether in the continuum limit diffeomorphism symmetry can be regained or not (this is related to the question whether the continuum limit can be performed or whether one remains with some 'fundamental discreteness' associated to the failure to fully satisfy the (pseudo) constraints) \cite{funddiscr}. \\
Having a canonical formalism with proper constraints which satisfy a first class algebra would completely circumvent this problem. Also one would use the diffeomorphism symmetry represented by the first class algebra to restrict hugely quantization (and discretization) ambiguities, see below. Such a formalism can be constructed from a discrete action which does display exact diffeomorphism invariance. Such actions (for non--topological theories, e.g. 4D gravity) will however be non--local\footnote{I.e. the couplings are not restricted to nearest neighbors. One can however expect an exponential decay with the lattice distance \cite{bietenholz}.} \cite{he}. This needs to be taken into account in the formulation of the canonical framework \cite{toappear}.

\item
{\it Discretization ambiguities:} Many different discretizations may lead to the same continuum limit. In other words, discretizations come usually with an overwhelming amount of ambiguities. This is not so much a problem if one sees the discretization just as a regularization of a continuum system. However, if one postulates discrete systems as fundamental, as some quantum gravity approaches do, one has to address the question of ambiguities. I.e. exclusion criteria should be formulated so that in the best case a unique theory can be found. \\
Requiring an exact realization of diffeomorphism invariance might provide a unique discretization, even on the quantum level. This has been proven for 1D reparametrization invariant (quantum) systems in \cite{seb}. The intuitive reason is the following: If there is a gauge symmetry that allows vertex displacements we can imagine to change the vertices such that there is a region where the effective lattice scale is macroscopic.  That is the discretization has actually to reproduce the macroscopic (continuum) physics (without any coarse graining taking place) -- all terms in the discrete action are therefore relevant. \\
Such a requirement can also be used to specify the path integral measure, as is shown in \cite{sebregge} for (linearized) Regge calculus. There one does actually require triangulation independence, which is however deeply related to diffeomorphism symmetry, as we will discuss in the next point.

\item
{\it Triangulation independence:} A discretization for which diffeomorphism symmetry is realized should also lead to triangulation independent results. I.e. expectation values or transition amplitudes should not depend on the choice of  (bulk) triangulation or lattice. Note that this means that one can go to the most coarse grained triangulation possible. Again, there is a proof for 1D (quantum) systems \cite{seb} which also gives an intuition why this should also hold in higher dimensions: If vertices of a triangulation can be displaced without interfering with predictions, we can also move vertices onto each other such that the triangulation is effectively coarse grained. \\
Examples for triangulation invariant quantum systems are well known from topological field theories, in particular 3D gravity with the Ponzano-Regge \cite{pr} and the Tuarev-Viro model \cite{tv}. Note however that it is the topological character (i.e. not having propagating degrees of freedom) of these theories that allows to have a triangulation invariant system with a partition function with only local  couplings. For interacting theories we have rather to expect non-local terms \cite{bietenholz, he}.

\item
{\it Continuum or large scale limit:} A discretization is usually adopted to describe the system on very small scales. To re-obtain physics on large scales, or on a continuum manifold as we know it, we have to take the continuum /large scale limit of the system. This can be done, for instance by coarse graining/renormalization, which will give effective actions describing physics on larger and larger scales. However from what has been said in the previous points, such a process is not really necessary for discretizations displaying diffeomorphism invariance. In particular we mentioned in the last point that such a theory should be triangulation independent, which would also include invariance under coarse graining. In other words, we are dealing with a system at a renormalization fixed point, see also the discussion in \cite{ditt}. As we will see constructing a diffeomorphism invariant discretization already includes the process of taking the continuum limit. In this sense the dichotomy of discretization and continuum symmetry is resolved.

\end{itemize}

In summary all advantages can be understood from the requirement for the discrete system to reproduce physics on all length scales, which entails that the discrete system already needs to encode continuum physics. Hence all the disadvantages coming with a discretization (in particular ambiguities, discretization dependence and consistency) can be addressed. 

On the other hand one needs to construct the discretization such that it can reproduce continuum physics. This requires a certain control over the solution space of the system. Indeed, one way to construct such a perfect discretization, is  -- turning the last point of our list on its head --by coarse graining and essentially finding the fixed point action \cite{wilson, hasenfratz}.  The process can be understood as `blocking from the continuum' \cite{bietenholz}, i.e. defining a lattice system which completely mirrors the physics of the continuum. There are a number of works where this approach has been successfully applied \cite{improve,proc,seb,he}. Concerning the question whether diffeomorphism symmetry can be regained in this way, the examples however include only systems where the perfect discretization is still local (i.e. topological systems or 1D). This is understandable due to the problem at hand: coarse graining basically means to solve the dynamics of the system. Therefore a perturbative approach is advisable. In \cite{he} the zeroth order of (free) field theories has been discussed including systems with gauge symmetries such as $U(1)$ gauge theory and gravity. Here one needs to carefully choose the coarse graining such that the linearized gauge freedom of these systems is preserved. 

In this work, using parametrized field theories as an example, we will discuss the challenges of going to higher order. The main point will be that, even before coarse graining, one has to make sure that the discrete action satisfies certain consistency requirements that hugely restrict the possible choices. These consistency requirements enforce the gauge symmetry to hold to the given order. This can of course be understood as an obstacle. On the other hand these conditions might allow the construction of a perfect action without  actually going through the coarse graining process completely. Furthermore an investigation of these consistency conditions might give us important information on the possible form of the perfect action, for instance regarding the structure of its non-local couplings. In particular the consistency conditions can be used to restrict possible discretization choices, with the option that there is only one unique  solution possible. Thus the consistency conditions allow the explorations of the perfect action without having it explicitly constructed yet.  From what has been said before the consistency conditions can be understood as infinitesimal versions for the requirement of triangulation independence of a given discretization or model. Thus these conditions could serve as the starting point for a systematic search for triangulation independent models.

\vspace{0.3cm}
We will discuss the issue of diffeomorphism symmetry in the discrete using as an example discretizations of parametrized (free) field theories, as in this case the split into physical and gauge variables is straightforward. In the next secion \ref{par} we will discuss the general features of such discrete theories and the basics of a perturbative expansion and coarse graining. We will discuss the coarse graining procedure order by order and see that at the first non-linear order consistency conditions arise. These have to be satisfied before coarse graining can be performed. Also the second order will be explicitly discussed as there new types of terms for the coarse grained action arise which do not appear in first order. We will give general formulas for the coarse graining of the different perturbative objects appearing in the expanded action. These results will be applied to the parametrized harmonic oscillator, which can be understood as $(0+1)$ dimensional parametrized field theory. Here we will see that the consistency conditions may be either violated already at first order or at second order. However a perturbative consistent discretization can be found and the perfect action can be constructed via coarse graining.  We will close with a discussion and outlook in section \ref{discuss}.

\section{Discretized  parametrized fields} \label{par}

Here we will discuss discretized parametrized (field) theories.  That is we assume a (here regular) lattice with vertices labelled by $x,y,\ldots$. Each of these vertices $x$ is described by embedding variables $t^a_x$ into $\Rl^d$, which is equipped with some (Euclidean or Minkowskian) metric. Furthermore we assume a (here scalar) field $f_x$ associated to the vertices.  The discrete theory is defined by the action
\ba\label{sat1}
S&=&  \frac{1}{2} S^{xy}(t) f_x f_y
\ea
where  we sum over repeated indices. We assume a quadratic action in the fields as we are dealing with a free field theory. This quadratic form depends on the induced lattice metric, which is reflected in the dependence of $S^{xy}$ on the embedding variables $t_z^a$ of the vertices $z$. It is important to note that both kind of variables, the fields $f_x$ and the embedding variables $t^a_x$ are treated as dynamical variables, i.e. we have to vary the action with respect to both fields. Changing the embedding variables will change the  matrix $S^{xy}(t)$ defining the quadratic interaction. This matrix includes metric information, i.e. the geometrical distance between the vertices (as the inverse defines the two point function of the theory). That is varying the embedding variables is actually a variation of the geometrical properties of the underlying lattice. If the position of the vertices is fixed by the equation of motions, it happens such that the action (\ref{sat1}) evaluated on the corresponding solutions for $f_x$ is extremal. i.e. such that the dependence of Hamilton's principal function on the embedding variables $t$ (now treated as parameters) is vanishing \cite{dittrichhoehn}.  In a sense the lattice itself (i.e. the vertex positions and therefore the geometric distances between vertices) is determined by the equations of motion for the embedding variables.

The discrete notion of diffeomorphism symmetry we are looking for would result in a gauge freedom for the variables $t^a_x$, i.e. independence of Hamilton's principal function from the vertex positions $t^a_w$. This entails an independence of physical predictions from the vertex positions and therefore the details of the lattice. Indeed such a feature can already be understood as discretization independence.

Diffeomorphism invariance is realized if given a solution $f_y^{\text{sol}}$ for the variables $f_y$
\ba
S^{xy}f^{\text{sol}}_y=0
\ea
the equation of motion associated to the $t_w^a$ are automatically satisfied (for arbitrary values of $t_w^a$)
\ba\label{em}
\frac{\partial S^{xy}}{\partial t^a_w} f^{\text{sol}}_x\,f^{\text{sol}}_y=0 \q  .
\ea
Here we basically have the derivative of the Lagrangian with respect to the metric, which defines the energy momentum. Indeed (\ref{em}) can be understood as a discrete conservation equation for the energy momentum \cite{marsden}. In case diffeomorphism symmetry is exact this conservation will hold for arbitrary vertex position. As one can easily see this is the case if  the derivative of $S^{xy}$ is of the form
\ba\label{form}
\frac{\partial S^{xy}}{\partial t^a_w}= \gamma^{wx}_{z,a}(t) S^{zy}(t)+ \gamma^{wy}_{z,a}(t) S^{zx}(t)
\ea
for some tensor field $\gamma^{wy}_{z,a}(t)$.
 In case the symmetry is broken, the requirement of energy conservation, i.e. equation (\ref{em}) will fix the vertex positions. Note however that we can linearize this theory around arbitrary vertex positions $t_w^a$ and the field configurations $f_x=0$. In this case the quadratic order of the action will be just given by the linearized field variables, the perturbations of the embedding variables will not appear. That is for the linearized theory we do have the gauge symmetry, as the linearized embedding variables will remain undetermined. This is analogous to the situation in Regge calculus \cite{roc,linearized}, where linearized diffeomorphism symmetry is realized around the flat solutions (based on arbitrary lattices). In the following we will see that continuing this gauge symmetry to higher order will require consistency conditions which are basically derived from (\ref{form}).

The framework we are discussing here should also be applicable to non-linear theories in particular Regge gravity. There all the variables are of the same form, namely given by the lengths of the lattice edges. That is a division into fields $f$ and embedding variables $t$ is not obvious. As it is not possible to solve the full theory at once, we will attempt a perturbative approach. Since in Regge calculus all variables are treated on an equal footing, we will attempt the same for parametrized field theory here, and expand both the fields $f$ and the embedding variables $t$  around a solution. We choose this solution to be the zero energy solution $f\equiv 0$ and $t$ determined by some regular lattice. That is we expand
\ba\label{sat2}
f &=& 0 + \epsilon \phi \nn\\
t^b &=&  n^b + \epsilon \xi^b \q .
\ea
Here $n^b$ is a vector in $\Zl^d$,
 denoting the regular lattice coordinates.  Using this expansion in the action (\ref{sat1}) we obtain 
\ba\label{s1}
\epsilon^{-2}S&=& \frac{1}{2}  \left( M^{xy}  \,\phi_x  \phi_y +  \eps  \,   \Gamma^{wxy}_a  \xi^a_w \,\phi_x \phi_y + \eps^2  \, \Gamma^{wzxy}_{ab}\xi^a_w\xi^b_z  \, \phi_x\phi_y + \ldots  \right)   \q . 
\ea
The unusual point in this perturbative expansion is that the variables $\xi$ do not appear in the quadratic term, but only start to appear in the cubic and higher order terms. This signifies that at lowest, zeroth, order, we will have gauge freedom as the lowest order term of the action does not depend on the variables $\xi$.  These variables do appear however at higher order and might (and indeed generically) lead to a breaking of gauge invariance. This will lead to severe conditions on the consistency of the perturbative expansion.

The aim here is to improve the discretization, i.e. the discrete action, towards better displaying the dynamics of the continuum theory. As in the continuum theory the (reparametrization) gauge symmetry is not broken, we might hope that in this way we obtain a discrete theory, in which this will be also the case. 

How can we obtain such an improvement? Here we will follow a renormalization/ coarse graining approach, i.e. we construct a family of (effective) actions on a given `coarse' lattice, that can be obtained by integrating out fine degrees of freedom from theories living on very fine lattices. These effective actions therefore display the dynamics defined on the finer lattice, and in an infinite refinement lattice, continuum dynamics. 

To this end we have to integrate out the fine grained degrees of freedom, i.e. solve the equations of motion on the fine lattice. This has to be done under the condition that the `microscopic' fields give under coarse graining the `macroscopic fields', which live on the coarse lattice. The solutions, that now depend on the coarse lattice, have to be re--inserted into the fine lattice action, which will result in an effective action depending on the macroscopic fields. 

This assumes a definition of coarse graining. Here (and also because of geometric considerations) we follow the simplest choice, equivalent to a decimation procedure.  That is the coarse grained fields on the coarse lattice site $X$ have (modulo a common factor) to coincide with the fine field at the lattice site $x=LX$ where $L$ denotes the lengths of a coarse graining block, that is \footnote{We could include a global scaling of the coarse grained variables, this global scaling can however easily inserted into the improved action terms at the end of the calculation.}
\ba
0&=& \Phi_X- \, \phi_{LX} \label{s1a}\\
0&=&  \Xi_X-\, \xi_{LX} \label{s1b} \q .
\ea
We will take care of these conditions by adding a Lagrange multiplier term 
\ba
\lambda^X \left( \Phi_X-\, \delta^x_{LX} \phi_x\right) \,+\, \alpha^X_a\left(\Xi^a_X- \delta^x_{LX} \,\xi^a_x\right)
\ea
thus introducing as further fields the Lagrange multipliers $\lambda^X$ and $\alpha^X_a$.

This finally defines the complete dynamical problem we have to consider. The equations of motion are obtained by varying the action (including Lagrange multiplier terms) with respect to the fields $\phi, \xi, \lambda, \alpha$ and are given by
\ba
0&=& M^{xy}\phi_y \,+\, \eps\, \Gamma_a^{wxy}   \xi^w_a \,  \phi_y  \,+\, \eps^2\,\Gamma_{ab}^{wzxy}   \xi^w_a \xi^z_b \, \phi_y  \,-\, \lambda^X \delta^x_{LX}  +\ldots \label{s2}  \\
0&=& \frac{1}{2} \eps \,\Gamma_a^{wxy}\phi_x\phi_y +\eps^2\, \Gamma_{ab}^{wzxy} \xi^b_z\, \phi_x\phi_y - \alpha^W_a \delta^w_{LW} +\ldots 
 \label{s3}  \\
0&=& \Phi_X- \phi_{LX} \label{s4}\\
0&=&  \Xi_X- \xi_{LX} \label{s5} \q .
\ea
where the dots signify higher order terms in $\epsilon$. As the equations of motions come naturally in orders of $\epsilon$ we will make a perturbative ansatz for the solutions
\ba\label{s6}
\phi_x&=& {}^0\!\phi_x+ \eps\,\,{}^1\!\phi_x+\ldots \nn\\
\xi_x&=& {}^0\!\xi_x+ \eps\,\,{}^1\xi_x+\ldots \nn\\
\lambda_X&=& {}^0\!\lambda_X+ \eps\,\,{}^1\!\lambda_X+\ldots \nn\\
\alpha^a_X&=& {}^0\!\alpha^a_X+ \eps\,\,{}^1\!\alpha^a_X+\ldots   \q .
\ea
(Here we will not expand the coarse grained fields $\Phi$ and $\Xi$, which are just parameters in the equations of motion. An expansion would just lead to more terms in the improved action which are however determined by the lower order terms, i.e. by expanding $\Phi$ and $\Xi$ in the improved action). Note that behind this perturbative ansatz there is an assumption, namely that the solutions can be actually expanded into a power series in $\epsilon$. We will see that this is generically not  the case for an action of the form (\ref{s1}). 

\subsection{Zeroth  order improvement}

Let us start with the zeroth order equation of motion whose solution will lead to the lowest order correction for the improved action. 

The zeroth order terms of the equations of motions  (\ref{s2})--(\ref{s5}) are given by
\ba
0&=& M^{xy}\zp_y-  \, \zl^X \delta^x_{LX} \label{s7}  \\
0&=& - \za^W_a \delta^w_{LW}
 \label{s8}  \\
0&=& \Zp_X- \zp_{LX} \label{s9}\\
0&=&  \Zx_X- \zx_{LX} \label{s10} \q .
\ea
Hence the $\zx_x$ with $x\neq LX$ remain undetermined. Furthermore we have $\za^W_a=0$ which leaves as with the coupled equations (\ref{s7},\ref{s9}). Assuming invertibility of $M^{xy}$ we can solve (\ref{s7}) for the fields $\zp_x$ and use this solution in (\ref{s9}) to determine the Lagrange multipliers $\zl^X$ as functions of the coarse fields $\Zp$. 
That is 
\ba
\zp_x &=&  \, (M^{-1})_{xy}   \delta^y_{LY}  \zl^Y \label{sat3}\\
\zl^X &=& {\cal M}^{XY} \Zp_Y \label{sat3b}
\ea
where ${\cal M}^{XY}$ is the inverse matrix to $({\cal M}^{-1})_{XY}:=(M^{-1})_{LX\,LY}$, which we also assume to be invertible. In this way we can write the solution as
\ba\label{s11}
\zp_x&=& \, (M^{-1})_{x(LX)} \, {\cal M}^{XY} \,\Zp_Y \,=:\, P_x^Y\Zp_Y  \q .
\ea
The map  $ P^Y_x:=(M^{-1})_{x(LX)} \, {\cal M}^{XY}$ provides us with the fine grained (zeroth order) solution for a given coarse grained configuration $\Phi_Y$. As we will see it will also appear for the coarse graining of the higher order terms.
 
These solutions (\ref{s11}) have to be inserted into the lowest order term of the  action to obtain the effective action ${}^0\! S'$, i.e. (we rescale by $\epsilon^{-2}$)
\ba\label{satu1}
{}^0\! S' \,&:=& \,  \frac{1}{2} M^{xy}  \zp_x \zp_y  \nn\\
&\underset{(\ref{s9},\ref{sat3})}{=}&   \frac{1}{2}  \Zp_X \,   \zl^X \nn\\
&\underset{(\ref{sat3b})}{=}&     \frac{1}{2}     \,  {\cal M}^{XY} \,  \Zp_X\Zp_Y %
\,\,=:\,\, \frac{1}{2}(M')^{XY}  \Zp_X   \,\Zp_Y  
\ea 
%
%
This is the improved action to zeroth order. Note that the improved quadratic form can be written as $(M')^{XY}={\cal M}^{XY}=P^X_x M^{xy}P^Y_y$.

We will assume that the refinement limit of (\ref{satu1}) exist and  will call the result the zeroth order perfect action.

\subsection{First order improvement}

We can now move on to the discussion of the next order. Two remarks are in order. Firstly we might use in  the expanded action (\ref{s1}) the zeroth order perfect action, i.e. replace $M_{xy}$ by $M'_{xy}$ there. In the following we will just denote both cases by $M_{xy}$. Secondly we will see that the first order improvement of the action does not strictly need the solutions of the first order equations of motion:
 to determine the first order term of the improved action insert the ansatz (\ref{s6}) into the action (\ref{s1}) and keep only terms up to first order in $\eps$. Note that  there are two types of terms in the first order improved action: The first type is of the form
\ba\label{s16}
\frac{\delta S}{\delta v}_{\,\,{\big|}v={}^0\!v}\,\cdot\,{}^1\!v
\ea
for the variables $v=\phi,\xi,\lambda,\alpha$. All these terms vanish, as ${}^0v$ satisfy the equations of motion for the zeroth order action and hence the first factor in (\ref{s16}) is zero.  The second type of terms comes from the $\eps^1$ term in  the action (\ref{s1})
which gives for the first order correction of the improved action
\ba\label{s17}
{}^1\!S'&=&
 \frac{1}{2} \, \Gamma^{wxy}_a \,     \zx^a_w \,   \zp_x\zp_y    \q .
\ea
Here the zeroth order variables $\zx^a_w$ appear, which remained however undetermined for $w\neq LW$. We should therefore check the first order equations of motions
\ba
0&=& M^{xy}\op_y+ \Gamma_a^{wxy} \zx^a_w \, \zp_y -  \ol^X \delta^x_{LX} \label{s12}  \\
0&=& \frac{1}{2} \Gamma_a^{wxy}\zp_x\zp_y  -\oa^W_a \delta^w_{LW}
 \label{s13}  \\
0&=& \op_{LX} \label{s14}\\
0&=&   \ox_{LX} \label{s15} \q .
\ea
Equations (\ref{s12}) and (\ref{s14}) can be solved similarly to the zeroth order equation and with the same assumptions, i.e. that $M^{xy}$ and  furthermore $({\cal M}^{-1})_{XY}:=(M^{-1})_{LX\,LY}$ can be inverted:
\ba\label{mon0}
\op_z&=&- \left[ (M^{-1})_{zv} \,-\, (M^{-1})_{z(LX)}{\cal M}^{X Y} (M^{-1})_{(LY)v}  \right] \, \Gamma^{wvy}_a\, \zx^a_w \zp_y     \q .
\ea

Equation (\ref{s13}) for $w \neq W$ is rather a consistency condition\footnote{This condition can be interpreted as energy momentum conservation.} as only the fields $\zp_x$ appear, which are already determined by the zeroth order equation. Hence we have to see this equation as a condition on the discretization, namely on $\Gamma_a^{wxy}$ and $M^{xy}$. Here we mention $M^{xy}$ as it determines the zero order solutions that appear in the conditions (\ref{s13}). 

Also the condition (\ref{s13}) ensures that the improved first order action does not depend on the choice of the undetermined variables $\zx_w^a$ for $w\neq LW$, as it can now be written as
\ba\label{s17a}
{}^1\!S'&=&
\,   \Zx^a_W \oa^W_a  \q .
\ea

Furthermore as (\ref{s13}) has to hold we can assume that $\Gamma$ has the following form
\ba\label{sgamma}
\Gamma^{wxy}_a= \gamma^{wx}_{a,z} M^{zy}+\gamma^{wy}_{a,z}   M^{zx} \q .
\ea
If we consider the case of not having any coarse graining conditions at all, i.e. $L=\infty$, the first order correction would be given as\footnote{The higher order solutions can be given only modulo solutions $\zp_x$ satisfying $M^{wx}\zp_x=0$. We assume that these terms can be put to zero by requiring appropriate boundary conditions.}
\ba
\op_z=-\gamma^{wy}_{a,z}\,\zx^a_w\,\zp_y  \q .
\ea
Requiring locality of the gauge action, that is a displacement of a vertex $w$ should only affect the field at the vertex $w$ we can conclude that $\gamma$ has to be of the form
\ba\label{sbeta}
\gamma^{wx}_{a,z}=\delta^w_z \beta^{wx}_a \q\q \text{so that} \q\q  \Gamma^{wxy}_a= \beta^{wx}_aM^{wy}+M^{wx} \beta^{wy}_a \q .
\ea
Note that we now may have indices appearing twice in upper positions, which we do not sum over. This form of $\Gamma$ ensures that the consistency conditions (\ref{s23}) are satisfied for $w\neq LW$.

Using the zeroth order solutions
\be
\zp_x = \, (M^{-1})_{xy} \,\delta^y_{LX}\,  {\cal M}^{XY} \,\Zp_Y   \,=\, P^Y_x\,\Zp_Y 
\ee
in (\ref{s17}) we can write the coarse grained action as
\ba
{}^1\!S' & =&  \frac{1}{2} \,\,
\Zx^a_W
  \left(  \beta^{(LW) x}_aP^X_x \,{\cal M}^{W Y} \,+\,  {\cal M}^{WX} \, \beta^{(LW)y}_aP^Y_y   \right)  \,\Zp_X      \,\Zp_Y  \nn\\
&=:& \frac{1}{2} \,\,\Zx^a_W
\left(    (\beta')^{WX}_{a} \, (M')^{WY}  \,\,+\,\,  (M')^{WX} (\beta')^{WY}_a \,\, \right) \,\Zp_X  \, \Zp_Y    \q .
\ea
This form of the improved $\Gamma'$ automatically ensures its consistency with respect to $M'$.
In summary, to find the coarse grained first order part of the action we have to construct
\ba\label{impg}
 (\beta')^{WX}_{a} \,\,=  \,\,
  \beta^{(LW)x}_{a} P^X_x\,=\,
  \beta^{(LW)x}_{a} (M^{-1})_{x (LY)} \,  {\cal M}^{YX}    \q .
\ea

Note that compared to the second order tensor $\Gamma^{wxy}_a$ we started originally with the improved tensor  might undergo two kinds of modifications. The first one might arise if we have to change $\Gamma$ such that the consistency conditions are satisfied (and the continuum limit still agrees with the continuum theory). The second modification is due to the actual coarse graining.

Furthermore if we require that the $\Gamma$ is of a consistent form and actually comes from the derivative of the action $S^{xy}(t)$ we will obtain rather strong conditions on $S^{xy}(t)$. Later--on we will discuss this issue for the 1D example.

\subsection{Second order improvement}

For the same reason for which we did not need the first order solutions to obtain the first order improvement of the action we will not need the second order solutions to obtain the second order improvement. 

 We will however need to consider the first order solutions and also have to check whether in the second order equations of motion consistency conditions arise. Again we might assume that the lower order tensors $M^{xy}$ and $\Gamma^{wxy}_a$ are the ones coming from a perfect discretization. 

The second order terms of the improved action are given by
\ba\label{m1}
{}^2\!S' &=&
\frac{1}{2}\bigg(
M^{xy}\op_x\op_y + \Gamma^{wxy}\zx^a_w\,(\zp_x\op_y+ \op_x\zp_y) 
+ \nn\\
&&\q\q\q\q\q\q\q\q\q\q\q   \q   \Gamma^{wxy}_a \ox^a_w \zp_z\zp_y   +
 \Gamma^{wzxy}_{ab} \zx^a_w \zx_z^b \zp_x\zp_y
\bigg)   \nn\\
&=&
\frac{1}{2}\bigg(
M^{xy}\op_x\op_y + \Gamma^{wxy}\zx^a_w\,(\zp_x\op_y+ \op_x\zp_y) 
+
 \Gamma^{wzxy}_{ab} \zx^a_w \zx_z^b \zp_x\zp_y
\bigg)  \q .
\ea
The first summand in the second line, which appears to depend on $\ox_w^a$,  does actually vanish due to the consistency conditions  (\ref{s13}) and the equations (\ref{s15}) which require $\ox_{LW}^a=0$. 
There is still the potential dependence on the variables $\zx^a_w$, which appear in (\ref{m1}), but which have not been determined so far by the zeroth and second order equations of motion. Note that also the solutions $\op_x$ depend on the variables $\zx^a_w$, see the  equations (\ref{mon0}).

Let us therefore investigate the second order part of the equations of motion
(\ref{s2})--(\ref{s5}) 
\ba
0&=& M^{xy}\,\sp_y+\Gamma_a^{wxy}   \left( 
\ox^a_w \zp_y  + \zx^a_w \op_y  \right) 
\,+\, 
\Gamma_{ab}^{wzxy}\, \zx^a_w \zx^b_z \, \zp_y - \, \sl^X \delta^x_{LX} \label{s20}\\
0&=& \frac{1}{2} \Gamma_a^{wxy} \left( \zp_x\op_y  + \op_x\zp_y \right)\, +\,   \zx^b_z \Gamma_{ab}^{wzxy}\zp_x\zp_y -\,\sa^W_a \delta^w_{LW}
 \label{s21}  \\
0&=& \sp_{LX} \label{s22}\\
0&=& \sx_{LX} \label{s23} \q .
\ea

Again equations (\ref{s21}) for $w\neq LW$ contain only variables of zeroth and first order. The embedding variables $\zx^a_w$ only appear in zeroth order. If we want to have gauge freedom with respect to these variables, we have to make sure that these consistency conditions are satisfied for arbitrary choices of $\zx^a_w$. Alternatively one can try to find solutions, which fix $\zx^a_w$ and coarse grain in this way. The hope however is that the resulting perfect action allows for full gauge freedom. To first find a discretization allowing for this gauge freedom to the order in question seems to be less cumbersome. This will also give the conditions that a discrete action has to satisfy, in order to display gauge freedom to second order.

We have to make some ansatz for the form of the tensor $\Gamma^{wzxy}_{ab}$. Here we use that it should arise as the second derivative of $S^{xy}(t^a_w)$ with respect to the embedding variables $t$. We know that the first order derivative (evaluated on some background) has the form (\ref{sbeta}). Another derivative acting on $M^{xy}=S^{xy}$ will produce a term of the same form. We therefore require 
\ba\label{Ga2}
\Gamma^{wzxy}_{ab}&=&
\frac{1}{2}\left(     \beta^{wy}_aM^{zw}\beta^{zx}_b\,+\,\beta^{wx}_aM^{zw}\beta^{zy}_b         \right) \,+ \nn \\
&&\frac{1}{2}\left(   \beta^{wy}_aM^{zx}\beta^{zw}_b\, +\, \beta^{wx}_a M^{zy}\beta^{zw}_b          \right)  \,+\,
\frac{1}{2} \left(       \beta^{zy}_aM^{wx}\beta^{wz}_b  \,+\, \beta^{zx}_aM^{wy}\beta^{wz}_b     \right)   \,+ \nn \\
&&\frac{1}{2} \left( M^{wx} \gamma^{wzy}_{ab} \,+\, \gamma^{wzx}_{ab}M^{wy} \right) \q  \,+\,\q
\frac{1}{2} \left( M^{zx} \gamma^{zwy}_{ab} \,+\, \gamma^{zwx}_{ab}M^{zy} \right)   \,+ \nn \\
&&\frac{1}{2} \left(  M^{wx}M^{zy} \eta^{wz}_{ab} \,+\, M^{zx}M^{wy} \eta^{zw}_{ab}  \right)  \q .
\ea
Here the terms in the first two lines arise through the derivative acting on the factor $M$ in $\Gamma=\beta M+ M \beta$. The other terms should be given by derivatives of $\beta$, with the terms in the third line suggested by the symmetry of second derivatives.   In the last line we just isolated\footnote{This splitting is not unique as the $\eta$ terms might be just reabsorbed into the $\gamma$ terms. However we will derive the condition $\gamma^{wzy}\zp_y=\delta^{wz}\gamma^{wzy}\zp_y$ for the part of $\gamma$ that does not vanish on solutions. This does not need to hold for $\eta^{wx}$. } terms, which might have two factors of $M$, from the terms in the fourth and fifth line. Note that this form of $\Gamma^{wzxy}_{ab}$ does not constitute a further requirement once we assume that the first order condition (\ref{sbeta}) holds for arbitrary background variables $t_w^a$ that is
\ba\label{w1}
\frac{\partial S^{xy}(t)}{t^a_w}&=& S^{wx}(t) \beta^{wy}_a(t) \,+\, \beta^{wx}_a(t) S^{wy}(t) \q .
\ea

Consider the second order solutions $\sp_x$ to (\ref{s20}) without coarse graining conditions, i.e. for $L\neq \infty$,
\ba\label{w2}
\sp_x&=&-\delta^w_x\beta^{wy}_a\ox_w^a \zp_y\,-\,\delta^w_x\gamma^{wzy}_{ab}\zx_w^a\zx_z^b \zp_y \q .
\ea
Requiring again locality of the gauge action, namely that displacing a vertex at $z\neq w$ should not affect the field at $w$, we can conclude that
\ba\label{w3}
\gamma^{wzy}_{ab}=\delta^{wz} \kappa^{wy}_{ab}  \q .
\ea

It will turn out that the  form (\ref{Ga2}) ensures that the consistency conditions (\ref{s21}) are satisfied for $w\neq LW$ and arbitrary values for $\zx_z^b$. 
Furthermore the solutions $\sa^W_a$ will not depend on $\zx_z^b$ with $z\neq LZ$. Indeed a  straightforward calculation gives
\ba\label{w4}
&& \frac{1}{2} \Gamma_a^{wxy} \left( \zp_x\op_y  + \op_x\zp_y \right)\, +\,   \zx^b_z \Gamma_{ab}^{wzxy}\zp_x\zp_y \q \;=\;\nn\\ 
&&
\delta^w_{LW}\bigg(
\frac{1}{2}\left(     (\beta')^{WY}_a{\cal M}^{ZW}(\beta')^{ZX}_b\,+\,(X \leftrightarrow Y)         \right) \,+ \nn \\
&&\q\q\;\,
\frac{1}{2}\left(   (\beta')^{WY}_a{\cal M}^{ZX}(\beta')^{ZW}_b\,+\,(X \leftrightarrow Y)          \right)  \,+ \,
\frac{1}{2} \left(       (\beta')^{ZY}_a{\cal M}^{WX}(\beta')^{WZ}_b \,+\,(X \leftrightarrow Y)      \right)   \,+ \nn \\
\, &&\q\q\;\,
\frac{1}{2} \left(  {\cal M}^{WX} \gamma^{(LW)(LZ)y}_{ab} (M^{-1})_{y(LV)} {\cal M}^{VY} \,+\,(X \leftrightarrow Y)    \right)   \,+\, 
\frac{1}{2} \Big(   \,\, W\leftrightarrow Z \,\,\Big)
 \,+ \nn \\
&&\q\q\;\,
\frac{1}{2}\left( {\cal M}^{WX} {\cal M}^{ZY}  \eta^{(LW)(LZ)}_{ab}  \,+\,(W \leftrightarrow Z)  \right) \,-\nn\\ 
&&\q\q\;\,
\frac{1}{2}\left( {\cal M}^{WX} {\cal M}^{ZY} \beta^{(LW)u}_a (M^{-1})_{uv} \beta^{(LZ)v}   \,+\,(W \leftrightarrow Z)  \right) \,+ \nn \\
&&\q\q\;\,
\frac{1}{2}\left( {\cal M}^{WX} {\cal M}^{ZY} \beta^{(LW)u}_a (M^{-1})_{u(LU)}   {\cal M}^{UV}     (M^{-1})_{(LV)v}       \beta^{(LZ)v}   \,+\,(W \leftrightarrow Z)  \right)
 \bigg)  \Xi_Z^a \Phi_X\Phi_Y  .\nn \\
\ea



Now we use for the first term in the improved action (\ref{m1}) the solutions (\ref{s12}) for $\op_x$ and find
\ba\label{m2}
M^{xy}\op_x\op_y
&=& - \op_v  \,\Gamma^{wvy}_a \zx^w_a\zp_y   \q .
\ea
Invoking the equations (\ref{s21}) the improved action can be rewritten as 
\ba\label{m3}
{}^2\!S' &=&  \frac{1}{2}\,\Zx^a_W\, \sa^W_a  \q 
\ea
where  due to the consistency equations (\ref{s21}) the solutions for $\sa^W_a$ are given by  (\ref{w4}).

Hence 
\ba\label{w5}
{}^2\!S' &=&  \frac{1}{2} (\Gamma')^{WZXY}_{ab}  \,\Zx^a_W \Zx^a_Z\,\Zp_X\Zp_Y 
\ea
where $\Gamma'$ is of the same form as $\Gamma$ in (\ref{Ga2}) with the replacements $M,\beta,\gamma,\eta \rightarrow M',\beta',\gamma',\eta'$ and the latter two tensors are given by
\ba\label{w6}
(\gamma')^{WZY}_{ab} &=&\gamma^{(LW)(LZ)y}_{ab} (M^{-1})_{y(LV)}{\cal M}^{VY} \; \,=\,  \gamma^{(LW)(LZ)y}_{ab}P^Y_y  \nn\\
(\eta')^{WZ}_{ab} &=& \eta^{(LW)(LZ)}_{ab} \,+\, \nn\\ &&
\frac{1}{2}\left( \beta^{(LW)u}_a  \left[ (M^{-1})_{u(LU)} {\cal M}^{UV}  (M^{-1})_{(LV)v}   \,- \,(M^{-1})_{uv}\right]      \beta^{(LZ)v } + (W\leftrightarrow Z ) \right).\q\q\q
\ea
Note that the ambiguity between $\gamma^{wzy}$ terms that include a $M^{wy}$ or $M^{zy}$ factor  and the $\eta^{wz}$ term does not matter, as both expressions will be coarse grained in the same way. Also note, that the expression
\ba
G_{xy}\,:=\, (M^{-1})_{xy}-(M^{-1})_{x(LX)}{\cal M}^{XY}(M^{-1})_{(LY)y}
\ea
is the Green's function associated to the matrix $M^{xy}$ together with the coarse graining conditions. That is, given the equations
\ba\label{jan0}
j^x\;=\;M^{xy}\phi_y-\delta^x_{LX}\lambda^X \q,\q\q \phi_{LX}=\Phi_X
\ea
the solution is given by
\ba\label{jan0b}
\phi_x=P^Y_x\Phi_Y\,+\, G_{xy}j^y  \q .
\ea
In summary, the coarse graining of the different objects appearing in the action will be determined by the two maps $P^Y_x$ and $G_{xy}$. This will be also the case at higher order.

The consistency requirements at higher order can be addressed in a similar fashion. The form of the higher order $\Gamma$ tensors can be obtained by taking the derivatives of the lower order ones and using the relations between the derivatives of $M$, $\beta$, etc..

\section{Discrete parametrized harmonic oscillator} \label{ossi}

Let us consider a simple but popular example \cite{proc,seb,ditt}, the discrete parametrized harmonic oscillator. We will start from a general family of actions, describing a parametrized $(0+1)$ dimensional free field $q_x$
%
%
\ba\label{ev1}
S= \frac{1}{2}\sum_{x\in \Zl}  \left[
D(t_{x+1}-t_x)\left(q_x^2+q_{x+1}^2\right) + 2 E(t_{x+1}-t_x) q_x q_{x+1} \right] \q .
\ea
A possible choice for the functions $D(t)$ and $E(t)$ is  
\ba\label{choi}
D(t)= \frac{1}{t}-m^2\alpha t  \q,\q\q E(t)= -\frac{1}{t}-\frac{1}{2}m^2(1-2\alpha) t
\ea
for instance with $\alpha=\tfrac{1}{2}$ or $\alpha=\tfrac{1}{4}$. 
Only the second choice will be perturbatively consistent to first order. 

For this example the perfect action is well known as it can be obtained from the continuum  Hamiltons principal function \cite{marsden,proc,seb}. It is given by
\ba\label{perf}
D=  m\frac{\cos(mt)}{\sin(mt)} \q,\q\q E= -m\frac{1}{\sin(mt)} \q . 
\ea

 Note that all these discrete actions lead to the same continuum limit for the solutions. The reason is that the coefficients $C^+$ and $C^-$ in (\ref{ev1}) in front of $Q_x^+=(q_x+q_{x+1})^2$ and of $Q_x^-=(q_{x}-q_{x+1})^2$ respectively, coincide in their lowest order expansion in $t$, as $C^+\sim -\frac{1}{4}m^2t$ and $C^-\sim \frac{1}{t}$.
   For solutions we will have $Q_x^-\sim t^2$, whereas $Q_x^2\sim t^0$, which explains the different scaling of the coefficients. 

To apply our formalism of improving the action order by order, we expand the variables as 
\ba\label{ev2}
q_x&=&0+\epsilon \phi_x \nn\\
t_x&=&a x +\eps \xi_x  
\ea
(here $x\in \Zl$) which results in an action of the form
\ba\label{ev3}
\epsilon^{-2}S&=& \frac{1}{2}  \left( M^{xy}  \,\phi_x  \phi_y +  \eps  \,   \Gamma^{wxy}  \xi_w \,\phi_x \phi_y + \eps^2  \, \Gamma^{wzxy}\xi_w\xi_z  \, \phi_x\phi_y + \ldots  \right)  \q .
\ea
We will now discuss the improvement of the different orders in (\ref{ev3}).

\subsection{Zeroth order}
We start with the zeroth order improvement for which we need the square term described by
\ba\label{ev4}
M^{xy} 
&=& 2 D \,\delta^{xy}  + E  \,  (\delta^{x (y+1)}  + \delta^{x (y-1)})   \q 
 \ea
where $D=D(a)$ and $E=E(a)$. 

The improved or perfect action to zeroth order can be calculated in many ways, one is as fixed points of the coarse graining flow \cite{seb}, the other is to explicitly obtain the inverse matrices $(M^{-1})_{xy}$ and then ${\cal M}^{XY}$ using Fourier transform as in \cite{he}. Here we will follow another route, by constructing explicitly the zeroth order solutions as functions of the coarse grained variables. Via the general form of the solution in (\ref{s11}) we will obtain the expression $(M^{-1})_{x (LX)} {\cal M}^{XY}$ that will also appear at higher order.

The homogeneous zeroth order equations of motion
\ba\label{ev5}
0&=&2D \zp_x +E \zp_{x-1}+ E\zp_{x+1}
\ea
can be easily solved with an ansatz $\zp_x=\exp(i \nu x)$  from which we obtain the condition
\ba\label{ev6}
\cos\nu &=& -\frac{D}{E}    \q ,
\ea
so that $\nu=ma$ for the perfect action (\ref{perf}). 
Choosing one of the roots for this equation we can write the general solution as
\ba
\zp_{LX+r}&=&A_X e^{i (LX+r)\nu} + B_X e^{-i (LX+r)\nu} 
\ea
where $r=0,\ldots,L-1$. Here, with making the coefficients $A_X,B_X$ dependent on the coarse grained intervals we indicate that the homogeneous equations of motions do not need to hold at the interval boundaries. Using the conditions $\zp_{LX}=\Zp_X$ and $\zp_{L(X+1)}=\Zp_{X+1}$ we can determine $A_X,B_X$ and obtain the solutions
\ba\label{e7}
\zp_{LX+r}&=& \frac{1}{\sin(\nu L)} \left[ \sin(\nu(L-r)) \,\Zp_X + \sin(\nu r) \,\Zp_{X+1} \right] \nn\\
&\underset{(\ref{s11})}{=}& (M^{-1})_{(LX+r)(LZ)}{\cal M}^{ZY} \Zp_Y \q .
\ea
We therefore have
\ba\label{e8}
 (M^{-1})_{(LX+r)(LZ)}{\cal M}^{ZY}  &=& \frac{1}{\sin(\nu L)} \left[ \sin(\nu(L-r)) \,\delta_{X}^Y + \sin(\nu r) \,\delta_{(X+1)}^Y \right]   \q .
\ea
With this at hand we can actually find easily the zeroth order improved action, as we just need  to multiply  (\ref{e8}) with the matrix $M^{(LW)(LX+r)}$ to find the matrix ${\cal M}^{WY}$ which defines the zeroth order improved action:
\ba\label{e9}
{\cal M}^{XY}= \frac{E \sin(\nu)}{\sin(L\nu)} \left[ -2  \cos(L\nu) \,\delta^{XY} + \delta^{X(Y-1)} +\delta^{X(Y+1)} \right] \q .
\ea
Here we also used the relation (\ref{ev6}) between the frequency $\nu$ and the parameters $D,E$.  In the continuum limit $a\rightarrow 0, L\rightarrow \infty $ such that $La=:a'=\text{const.}$ we obtain for any of the choices (\ref{choi},\ref{perf}) the perfect action
\ba\label{e9p}
{\cal M}^{XY} &=& 2\frac{m\cos(ma')}{\sin(ma')}\,\delta^{XY}   - \frac{m}{\sin(ma')} \left( \delta^{X(Y-1)} +\delta^{X(Y+1)} \right) 
\ea
on the coarse grained lattice.

\subsection{First order}

To find the tensor $\Gamma^{wxy}$ appearing in the expanded action (\ref{ev3}) we take the derivative of
\ba\label{dec01}
S^{xy}=  D(t_{x+1}-t_x)\delta^{xy}+D(t_{x}-t_{x-1})\delta^{xy} 
+ E(t_{x+1}-t_x)\delta^{x(y-1)}+ E(t_{x}-t_{x-1})\delta^{x(y+1)} \,
\ea
with respect to $t_w$  (here a prime will denote the derivative of a function)
\ba\label{dec03}
\frac{\partial}{\partial t_w} S^{xy}&=&
D'(t_{x+1}-t_x)\delta^{xy} \left[ \delta^{w(x+1)}-\delta^{wx} \right] \;+\; D'(t_{x}-t_{x-1})\delta^{xy} \left[ \delta^{wx}-\delta^{w(x-1)} \right] +\nn\\
&& E'(t_{x+1}-t_x)\delta^{x(y-1)}\left[ \delta^{w(x+1)}-\delta^{wx} \right] + E'(t_{x}-t_{x-1})\delta^{x(y+1)} \left[ \delta^{wx}-\delta^{w(x-1)} \right] \q .\q\q\q
\ea
Putting $t_{x+1}-t_x=a$ for all $x$ we can write $\Gamma^{wxy}$ as
\ba\label{dec04}
\Gamma^{wxy}&=& E'\delta^{wx}\left[ \delta^{w(y+1)}-\delta^{w(y-1)}  \right]+  E'\delta^{wy}\left[ \delta^{w(x+1)}-\delta^{w(x-1)}  \right] +\nn\\
&&
\tfrac{1}{2} D'\delta^{w(x+1)}\left[ \delta^{w(y+1)}-\delta^{w(y-1)}  \right]
+\tfrac{1}{2} D'\delta^{w(y+1)}\left[ \delta^{w(x+1)}-\delta^{w(x-1)}  \right]   + \nn\\
&&
\tfrac{1}{2} D'\delta^{w(x-1)}\left[ \delta^{w(y+1)}-\delta^{w(y-1)}  \right]
+\tfrac{1}{2} D'\delta^{w(y-1)}\left[ \delta^{w(x+1)}-\delta^{w(x-1)}  \right]   \nn\\
&=& \tilde M^{wx} \left[ \delta^{w(y+1)}-\delta^{w(y-1)}  \right]+ \tilde M^{wy} \left[ \delta^{w(x+1)}-\delta^{w(x-1)}  \right]  \q .
\ea
Note that in the second line we just subtracted two terms  which are added again in the third line. Here $\tilde M^{xy}$ is given by\ba\label{dec05}
\tilde M^{xy} &=&  E' \delta^{xy}+\tfrac{1}{2}D' \delta^{x(y+1)} + \tfrac{1}{2}D' \delta^{x(y-1)} \q .
\ea
Hence we can satisfy the consistency requirement  (\ref{sbeta}) if $\tilde M^{xy} =\beta M^{xy}$. That is we obtain the conditions
\ba\label{condis1}
D'(a)\;=\;2\beta(a) E(a)\q,\q\q E'(a)\;=\;2\beta(a) D(a)  \q\Rightarrow\q   DD'\;=\; EE' \;=\;2\beta E D\q .
\ea
From the family of actions (\ref{choi}) only the choice with $\alpha=\tfrac{1}{4}$ satisfies (\ref{condis1}) with $\beta(a)=\tfrac{1}{2a}$. For the perfect action (\ref{perf}) the condition is also satisfied with $\beta=-\tfrac{E}{2}$.

In general we can make a ansatz for $E,D$ in (odd) powers of $a$
\ba\label{ans1}
D(a)&=&\frac{1}{a}+d_1 a +d_2 a^2 +\ldots \nn\\
E(a)&=&-\frac{1}{a}+e_1 a + e_2 a^2+\ldots  \q .
\ea
Here the coefficients of $a^{-1}$ are determined by the continuum limit. With this form we can conclude that $\beta(a)=\frac{1}{2a}+O(a)$.

A consistent form of $\Gamma^{wxy}$ is therefore given by
\ba
\Gamma^{wxy}=\beta^{wx}M^{wy}+M^{wx}\beta^{wy}
\ea
where
\ba
\beta^{wx}  \;=\; \beta(a) \left[ \delta^{w(x+1)}-\delta^{w(x-1)}  \right]  \q  .
\ea 
with $\beta(a)=\frac{1}{2a}+O(a)$.

%


Following (\ref{impg}) it is straightforward to determined the improved $(\beta')^{WX}$ to  
\ba\label{e12}
(\beta')^{WX}&=&  
\beta^{(LW)y}(M^{-1})_{y(LZ)}{\cal M}^{ZX} \nn \\
&=& \frac{\beta(a)\sin(\nu)}{\sin(L\nu)}  \left[ \delta^{W(X+1)}-\delta^{W(X-1)}  \right] \nn\\
& \underset{\footnotesize a\rightarrow 0,L\rightarrow \infty}{\longrightarrow} &\frac{m}{2\sin(a'm)}   \left[ \delta^{W(X+1)}-\delta^{W(X-1)}  \right]  \q .
\ea
Notice that for taking the continuum limit in the last line it is sufficient to know that $\beta(a)=\frac{1}{2a}+O(a)$.
For all such choices we re-obtain the first order of the perfect action.

\subsection{Second order}

In (\ref{Ga2}) we determined a consistent form of the second order tensor $\Gamma^{wzxy}$ 
\ba\label{Ga2h}
\Gamma^{wzxy}_{\text{cons}}&=&
\frac{1}{2}\left(     \beta^{wy}M^{zw}\beta^{zx}\,+\,\beta^{wx}M^{zw}\beta^{zy}         \right) \,+ \nn \\
&&\frac{1}{2}\left(   \beta^{wy}M^{zx}\beta^{zw}\, +\, \beta^{wx} M^{zy}\beta^{zw}       \right)  \,+\,
\frac{1}{2} \left(       \beta^{zy}M^{wx}\beta^{wz}  \,+\, \beta^{zx}M^{wy}\beta^{wz}     \right)   \,+ \nn \\
&&\frac{1}{2} \left( M^{wx} \gamma^{wzy} \,+\, \gamma^{wzx}M^{wy} \right) \q  \,+\,\q
\frac{1}{2} \left( M^{zx} \gamma^{zwy} \,+\, \gamma^{zwx}M^{zy} \right)   \,+ \nn \\
&&\frac{1}{2} \left(  M^{wx}M^{zy} \eta^{wz} \,+\, M^{zx}M^{wy} \eta^{zw}  \right)  \q .
\ea
Note that we will assume $\gamma^{wzy}\sim\delta^{wz}$ as this ensures locality of the gauge action.
On the other hand the second order derivative of the second rank tensor (\ref{dec01}) gives
\ba\label{dec06}
\Gamma^{wzxy}= \frac{1}{2}\frac{\partial^2}{\partial t_z\partial t_w} S^{xy}&=&
\frac{1}{2} D''(t_{x+1}-t_x)\delta^{xy} \left[ \delta^{w(x+1)}-\delta^{wx} \right] \left[ \delta^{z(x+1)}-\delta^{zx} \right] \;+\;
\nn\\
&&\frac{1}{2} D''(t_{x}-t_{x-1})\delta^{xy} \left[ \delta^{wx}-\delta^{w(x-1)} \right]        \left[ \delta^{zx}-\delta^{z(x-1)} \right]         \;+\;\nn\\
&& \frac{1}{2}E''(t_{x+1}-t_x)\delta^{x(y-1)}\left[ \delta^{w(x+1)}-\delta^{wx} \right] \left[ \delta^{z(x+1)}-\delta^{zx} \right] \;+\;\nn\\
&&\frac{1}{2} E''(t_{x}-t_{x-1})\delta^{x(y+1)} \left[ \delta^{wx}-\delta^{w(x-1)} \right] \left[ \delta^{zx}-\delta^{z(x-1)} \right] \; .
\ea

We can compare these two expressions for different values of the indices $(wzxy)$. This will give a number of equations involving $E,D$, its second derivatives, and the components of $\beta, \gamma, \eta$. In addition we have the conditions (\ref{condis1}) from our discussion of the first order. 


For instance combining the equations for $(wzxy)=(0011)$ and $(1120)$ on the one hand and for $(0101)$ and $(1230)$ on the other, we find the conditions
\ba\label{condis2a}
8D\beta^2\,=\,D''\q,\q \text{and}\q 4\beta^2E^2 + 4 \beta^2 D^2=-EE'' \q .
\ea
Here, $\beta$ can be also expressed as a quotient between $D'$ and $E$ or between $E'$ and $D$ from (\ref{condis1}), so that
\ba\label{system}
2\frac{(E')^2}{D}\,=\,D''\,=\,2\frac{D(D')^2}{E^2}\q, \q \text{and}\q  (D')^2+(E')^2=-EE''  \q .
\ea
None of the actions (\ref{choi}) satisfies these requirements, that is none of these is perturbatively consistent to second order.  Of course the perfect action (\ref{perf}) does satisfy (\ref{condis2a}). Hence we can regain the perfect action by solving the system (\ref{system}) via an ansatz for $D,E$ as a power series in $a$ (starting with $a^{-1}$). 

From the equations for $(wzxy)=(1230),(1120),(1121),(1101),(1111)$ we can obtain conditions for the components of $\gamma$ and $\eta$
\ba\label{condis2b}
\eta^{12} &=& -\frac{\beta^2}{E} \q\q\q\q\;=\;\eta^{21} \nn\\
  \gamma^{111}&=& \frac{D''}{4D} - D \eta^{11}\nn\\
 \gamma^{110}&=&\beta^2\frac{D}{E}-\frac{1}{2}E\eta^{11}\;=\;\gamma^{112} \q .
\ea
This fixes the off--diagonal elements of $\eta$ but leaves an ambiguity between the diagonal elements of $\eta$ and the components of $\gamma$. However this ambiguity is inherent in our definition (\ref{Ga2h}) as the diagonal elements of $\eta$ can be reabsorbed into an additional term of the $\gamma$ tensor, which is proportional to $M$. (This only applies to the diagonal elements of $\eta$ as we require $\gamma^{wzy}\sim\delta^{wz}$.) 
This possibility of redefining $\gamma$ allows to set the components $\gamma^{110},\gamma^{112}$ to zero.  With this convention we obtain
\ba
\gamma^{wzy}&=&  2\beta^2\left(1-\frac{D^2}{E^2}\right) \nn\\ &=&
 \frac{m^2}{2} \delta^{wz}\delta^{wy}+O(a^2) \nn\\
\eta^{wz}&=& 2\beta^2\frac{D}{E^2} \delta^{wz} -\frac{\beta^2}{E}\left(\delta^{w(z+1)} +\delta^{w(z-1)} \right) \nn\\&=&
\frac{1}{a}\left(  \frac{1}{2} \delta^{wz} + \frac{1}{4}\delta^{w(z+1)} +\frac{1}{4}\delta^{w(z-1)} \right)  \,\,+O(a)\,\,\q .
\ea
In the second and fourth line we have given the lowest orders in $a$ of the $\gamma$ and $\eta$ tensor respectively. These can be easily obtained from the information about the lowest orders of $D,E$ and $\beta$ we collected so far and will be sufficient for the purpose of coarse graining from the continuum.

Following (\ref{w6}) let us now perform the coarse graining
\ba\label{jan1}
(\gamma')^{WZY} &=&\gamma^{(LW)(LZ)y} (M^{-1})_{y(LV)}{\cal M}^{VY} \; \,=\,  \gamma^{(LW)(LZ)y}P^Y_y  \nn\\
(\eta')^{WZ} &=& \eta^{(LW)(LZ)}\,+\, \nn\\ &&
\frac{1}{2}\left( \beta^{(LW)u}  \left[ (M^{-1})_{u(LU)} {\cal M}^{UV}  (M^{-1})_{(LV)v}   \,- \,(M^{-1})_{uv}\right]      \beta^{(LZ)v } + (W\leftrightarrow Z ) \right) \nn\\
&=&  \eta^{(LW)(LZ)} -\frac{1}{2}\beta^{(LW)u} G_{uv}   \beta^{(LZ)v } - (W \leftrightarrow Z) \q .
\ea
of these tensors. To this end we will need the Green's function $G_{uv}$. It can be obtained for instance by Fourier transform or by solving the system of equations (\ref{jan0}) directly and by comparing the solution to (\ref{jan0b}). Such a solution can be constructed by mimicking the variation of constants method from the continuum. One will obtain
\ba\label{jan2}
G_{(LX+r)(LY+s)}&=& \delta_{XY}\bigg( \,  \frac{\sin(\nu r) \sin(\nu s)}{E\sin(\nu)} \frac{\cos(L\nu)}{\sin(L\nu)}  \q -\q  \frac{\sin(\nu r) \sin(\nu s)}{E\sin(\nu)} \delta_{rs}  \nn\\
&&\q\; \;  - \frac{\sin(\nu r) \cos(\nu s)}{E\sin(\nu)} \,\theta(s-r) \;-\;  \frac{\sin(\nu s) \cos(\nu r)}{E\sin(\nu)} \,\theta(r-s) \bigg) \q 
\ea
where
\ba
\theta(r) =
\begin{cases} 0 & \text{if $r\leq 0$,}
\\
1 &\text{if $r > 0$.}
\end{cases}
\ea
We can now compute the coarse grained entities $\gamma',\eta'$ and obtain in the limit $a\rightarrow 0,L\rightarrow \infty$ with $L\cdot a=a'$
\ba
(\gamma')^{WZY} &=& \frac{1}{2}m^2  \delta^{WZ} \delta^{WY} \nn\\
(\eta')^{WZ}&=& \frac{m}{4 \sin(a' m )} \left(  \delta^{W(Z+1)}+ \delta^{W(Z-1)}   \right) +\frac{m}{2} \frac{\cos(a' m)}{\sin(a' m)} \delta^{WZ} \q .
\ea
As expected this result coincides with the tensors $\gamma_{perf},\eta_{perf}$ coming from the perfect action.

\subsection{Summary}

We have seen that from the family (\ref{choi}) of discrete actions only the choice $\alpha=\tfrac{1}{4}$ is perturbatively consistent to first order. None of these choices is however perturbatively consistent to second order.  Nevertheless we succeeded to find a perturbatively consistent second order term. Note that here different strategies are possible: we only determined the altered second order term to the lowest order in the lattice constant $a$ as the perfect action could afterwards be constructed through coarse graining. Alternatively one might demand that the second order term arises from the second derivate of the action, that is that the differential equations (\ref{condis2a},\ref{system}) are satisfied. This would also fix the higher order terms in the lattice constant $a$ so that even without coarse graining, the perfect action can be calculated from the solutions of these differential equations.

\section{Discussion} \label{discuss}

The notion of diffeomorphism symmetry in the discrete advocated here is a very powerful one: it leads to discretization independence and  to a reconciliation between continuum space time  and discrete underlying lattice. Basically such a symmetry requires that the discretization encodes already continuum physics. It should therefore not be very surprising that there are no examples yet where such a symmetry is realized for proper field theories with propagating degrees of freedom. Generically discretizations will rather feature broken diffeomorphism symmetry, in the sense that there exist special background solutions for which the symmetry is realized, but is violated if considering other solutions. These special solutions are often the most simple ones around which a perturbative expansion would be natural. 

A theory with broken gauge symmetries is however not perturbatively consistent if expanded around a background solution where the gauge symmetry is realized. For Regge calculus this applies to the expansion around flat space (or for Regge calculus with cosmological constant and curved tetrahedra \cite{newregge} around homogeneous background solutions). That is one should be aware that, i.e. the scattering of gravitons on a lattice is a priori not well defined.  However the computation of such scattering amplitudes provides an important test for the low energy behavior of quantum gravity theories such as spin foams \cite{spezialereview}.

One possibility to allow for a perturbative treatment of lattice action with broken gauge symmetries is to change the discretization to the appropriate order. This change should ensure that the gauge symmetry is realized to this order, allowing a consistent perturbative treatment. That is gauge modes will decouple to the given order and only physical modes need to be considered.

The change of discretization is already an important step towards constructing a perfect action, i.e. a discretization where the gauge symmetry in question -- diffeomorphism symmetry -- is fully realized. It might therefore be quite non--trivial to find perturbatively consistent discretizations for field theories with propagating degrees of freedom. Future investigations will show, whether such perturbatively consistent discretizations require non-local couplings. Such couplings have to be expected for the perfect action as they do appear under coarse graining. In this case the consistency requirements can give important informations on the structure of the perfect action, without having explicitly constructed it yet.

In 4D Regge calculus difffeomorphism symmetry is broken -- in general to quadratic order \cite{bahrdittrich1}. This does not exclude the possibility that Regge calculus is perturbatively consistent to first non--linear order on a regular lattice, which would make the regular lattice a preferred choice. This can be checked explicitely, as the conditions for consistency are analogous to those for parametrized field theory.

Here we discussed the classical theory, i.e. tree level amplitudes. A much farther reaching question is to generalize the considerations to quantum theory (see \cite{seb} for a discussion of the quantum theory for the discrete anharmonic oscillator). There are a number of crucial questions to consider. In particular the minimal distance on a lattice serves usually as a regulator. However if vertex translation symmetry is realized such a minimal distance looses its meaning as vertices can be moved on top of each other. That is, in such a lattice quantum theory with diffeomorphism symmetry the lattice looses is function as a regulator and finiteness needs to be provided for by other means. One possibility is through the discreteness of spectra of geometric operators, which is realized in loop quantum gravity \cite{disc}. Insight into this issue can be gained by finding the fixed points under coarse graining of quantum gravity models, such as spin foams \cite{oeckl,ecki}

\section*{Acknowledgements}
It is a pleasure to thank the organizers of the 3rd Quantum Gravity and Quantum Geometry School for the kind invitation  and Benjamin Bahr as well as Ralf Banisch for collaboration. Research at Perimeter Institute is supported by the Government of Canada through Industry Canada and by
the Province of Ontario through the Ministry of Research and Innovation.

\vspace{1cm}



\end{document}